\documentclass[conference]{IEEEtran}
\IEEEoverridecommandlockouts
\usepackage{cite}
\usepackage{amsmath,amssymb,amsfonts}
\usepackage{algorithmic}
\usepackage{graphicx}
\usepackage{tabularx}
\usepackage{textcomp}
\usepackage{xcolor}
\usepackage{stfloats}
\usepackage{multirow}
\usepackage{makecell}
\usepackage{subfigure}
\def\BibTeX{{\rm B\kern-.05em{\sc i\kern-.025em b}\kern-.08em
    T\kern-.1667em\lower.7ex\hbox{E}\kern-.125emX}}
\begin{document}

\title{A 2D Non-Stationary Channel Model for Underwater Acoustic Communication Systems}
\author{Xiuming Zhu\textsuperscript{1,2}, Cheng-Xiang Wang\textsuperscript{1,2,*}, Ruofei Ma\textsuperscript{3}
	\\
	\textsuperscript{1}National Mobile Communications Research Laboratory, School of Information Science and Engineering,
	\\Southeast University, Nanjing 210096, China.\\
	\textsuperscript{2}Purple Mountain Laboratories, Nanjing 211111, China.\\
	\textsuperscript{3}School of Information Science and Engineering, Harbin Institute of Technology, Weihai 264209, China.
	\\
	\textsuperscript{*}Corresponding Author: Cheng-Xiang Wang
	\\
	Email: \{xm\_zhu, chxwang\}@seu.edu.cn, ruofeimahit@gmail.com

}

\maketitle

\begin{abstract}
	Underwater acoustic (UWA) communication plays a key role in the process of exploring and studying the ocean.
	In this paper, a modified non-stationary wideband channel model for UWA communication in shallow water scenarios is proposed.
	In this geometry-based stochastic model (GBSM), multiple motion effects, time-varying angles, distances, clusters' locations with the channel geometry,
	and the ultra-wideband property are considered, which makes the proposed model more realistic and capable of supporting long time/distance simulations.
	Some key statistical properties are investigated, including temporal autocorrelation function (ACF), power delay profile (PDP), average delay,
	and root mean square (RMS) delay spread.
	The impacts of multiple motion factors on temporal ACFs are analyzed.
	Simulation results show that the proposed model can mimic the non-stationarity of UWA channels.
	Finally, the proposed model is validated with measurement data.
\end{abstract}

\begin{IEEEkeywords}
	Underwater acoustic communication, shallow water, channel modeling, GBSM, non-stationarity
\end{IEEEkeywords}

\section{Introduction}
As a potential technology in supporting underwater communications in the sixth generation (6G) space-air-ground-sea integrated networks\cite{wang2020_6g,you2020_6g},
UWA communication plays a key role in exploring and studying the ocean.
Since channel models are essential for the design and evaluation of communication systems\cite{wang2018survey}, accurate UWA channel models with low complexity and good flexibility are indispensable.

Due to the complex propagation environments, UWA channels show several unique characteristics.
UWA channels are affected by numerous motion factors\cite{qarabaqi2013,baktash2015}, which can be divided into:
1) intentional platform motion, e.g., autonomous underwater vehicle's (AUV's) vehicular motion;
2) unintentional drifting platform motion caused by the movement of water; 3) surface motion, which may fluctuate with waves and show cyclic patterns\cite{rudander2017}.
Because of the low speed of sound in the water (usually about 1500 m/s), the changes of transmission delays caused by motion effects cannot be neglected\cite{stojanovic2009,vanwalree2013}.
Moreover, UWA channels may exhibit non-stationarity as a result of time-varying delays and ultra-wideband property, violating the wide sense stationary uncorrelated scattering (WSSUS) assumption\cite{vanwalree2013,vanwalree2013a}.


A number of studies have worked on the UWA communication channel modeling\cite{gul2017,huang2019,qarabaqi2013,socheleau2010,baktash2015,naderi2017,naderi2018,zajic2011}.
However, no standardized channel model has been proposed yet\cite{song2019}.
In \cite{gul2017,huang2019}, typical solutions were used to calculate the acoustic field, e.g., ray tracing and parabolic approximation.
However, these models, while accurate, are deterministic and lack of flexibility.
In \cite{qarabaqi2013,socheleau2010,baktash2015,naderi2017,naderi2018,zajic2011}, several stochastic channel models were proposed to characterize UWA channels.
In these models, GBSMs have good balance among accuracy, complexity, and flexibility\cite{naderi2017,naderi2018,zajic2011}, and are widely used in the modeling of wireless communication channels.
In \cite{naderi2017,naderi2018}, GBSMs based on WSSUS assumption were proposed and analyzed, which cannot capture the non-stationarity of UWA channels.
In \cite{zajic2011}, the effects of platform and scatterers' motion were considered.
However, the velocities of the transmitter (Tx) and receiver (Rx) were assumed to be constant, which could not model the random drifting motion in long time simulations.
The delay changes caused by surface motion were modeled as Gaussian processes, which cannot characterize the cyclic patterns in channel properties.
Moreover, the model neglected the change of the channel geometry due to significant displacements of Tx and Rx,
thus making the model not suitable for long time/distance simulations, e.g., AUV's application scenarios.

To the best of the authors' knowledge, non-stationary GBSMs for shallow water scenarios considering multiple motion factors and allowing for long time/distance communication scenarios are still missing in the literature.
This work is aiming to fill the research gap.
The major contributions and novelties of this work can be summarized as follows.
\begin{enumerate}
	\item {The proposed model considers two kinds of platform motion (constant intentional motion and random drifting motion) and sinusoidal motion of surface scatterers.
	      The distances, angles, especially the clusters' locations are modeled as time-varying parameters.}
	\item {The time-frequency varying transmisson losses caused by time-varying propagation distances and underwater frequency-dependent absorption loss are considered in the proposed model.}
	\item {The proposed twin-cluster GBSM supports single-bounce (SB) and multiple-bounce (MB) propagation. For MB propagation, the angles-of-arrival (AoAs) are independent with the angles-of-departure (AoDs), while these angles are correlated with geometric relationships for SB propagation.}
	\item {Based on the proposed model, some important statistical properties such as ACF, PDP, average delay, and RMS delay spread are studied and analyzed. The proposed GBSM is also validated with measurement data.}
\end{enumerate}

The rest of this paper is organized as follows.
In Section~\uppercase\expandafter{\romannumeral2}, a modified non-stationary shallow water GBSM is introduced in detail.
The key statistical properties of the model are derived in Section~\uppercase\expandafter{\romannumeral3}.
Section~\uppercase\expandafter{\romannumeral4} presents the simulation results and analysis.
Conclusions are finally drawn in Section~\uppercase\expandafter{\romannumeral5}.

\section{A Modified Non-stationary GBSM for Shallow Water UWA Communication}
\begin{figure}[t]
	\centerline{\includegraphics[width=0.42\textwidth]{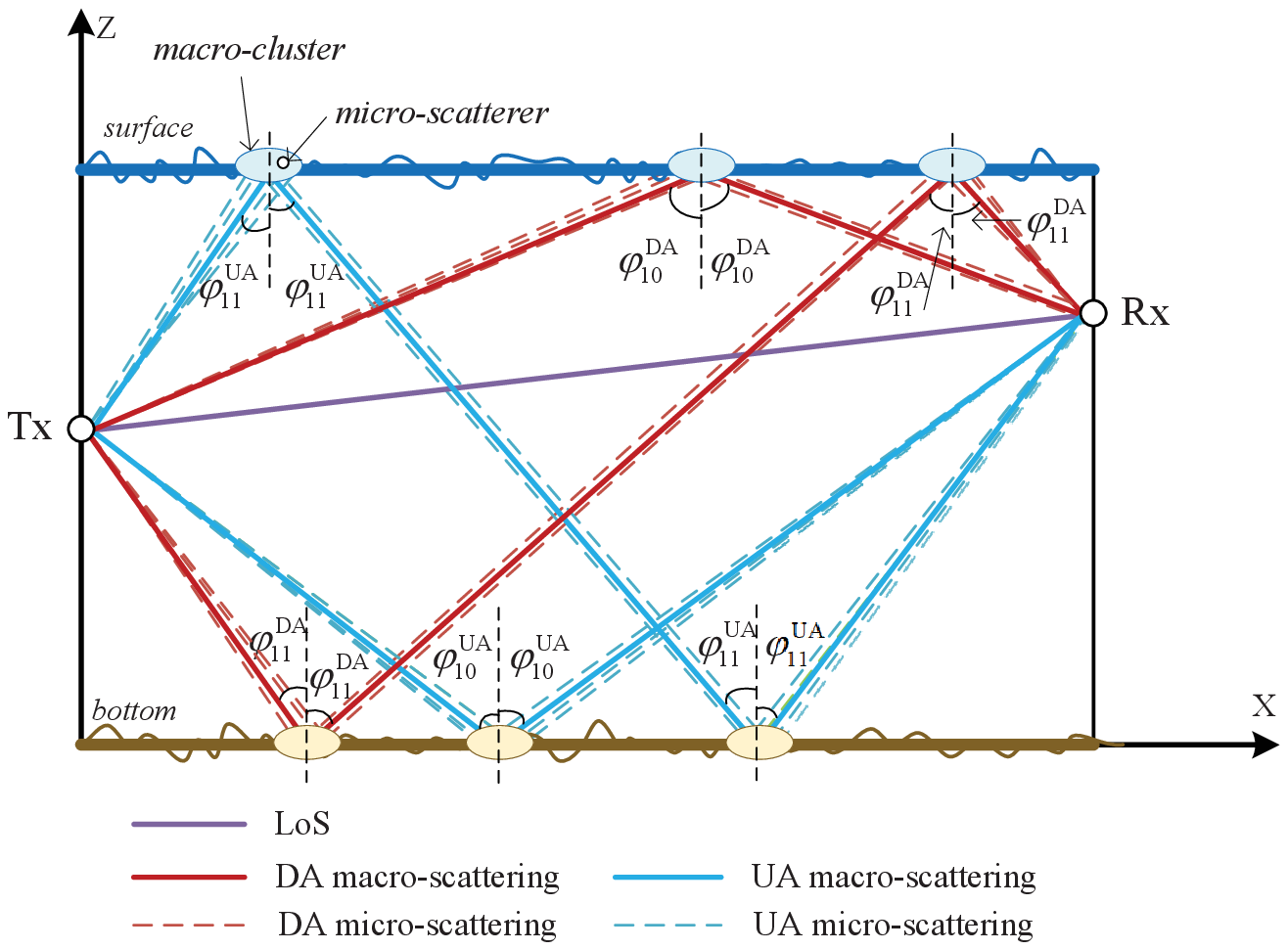}}
	\caption{Illustration of different propagation components ($N_S=1, N_B=1$).}
	\label{fig_1}
\end{figure}

Let us consider a single-input single-output (SISO) communication system in a shallow water scenario, where both Tx and Rx are located in the water.
The sound speed in the water is related to salinity, temperature, and pressure. In shallow water scenarios, these factors are usually constant because of the mixing of wind, so is the sound speed ($\approx 1500$ m/s)\cite{stojanovic2009,zhou2014}. Thus, acoustic signals can be assumed to propagate along straight lines in shallow water scenarios.
The acoustic signals may reach the Rx directly or reflect several times between two boundaries.
Fig. \ref{fig_1} gives an illustration of different propagation components, i.e., line-of-sight (LoS), downward arrival (DA), and upward arrival (UA) paths.
Due to the roughness of the boundaries, it is assumed that each DA (UA) path is comprised of $N_{s \tilde{b}}^{\rm DA}$ ($M_{b \tilde{s}}^{\rm UA}$) diffuse scattering rays.
The specular reflection ray is called the macro-scattering ray, while the diffuse rays are defined as micro-scattering rays \cite{naderi2017}.
The micro-scatterers are assumed to be clustered around the macro-cluster, i.e., the specular reflection point.
Each macro-scattering ray is assumed to be the average of micro-scattering rays.
Assuming that each DA (UA) path will contact the surface (bottom) at most $N_S$ ($N_B$) time(s) with $s$ ($\tilde{s}$) surface interaction(s) and $\tilde{b}$ ($b$) bottom interaction(s),
then there are constraints $1 \leq s \leq N_{S}$ ($1 \leq b \leq N_B$) and $s-1 \leq \tilde{b} \leq s$ ($b-1 \leq \tilde{s} \leq b$).

\begin{figure}[t]
	\centerline{\includegraphics[width=0.42\textwidth]{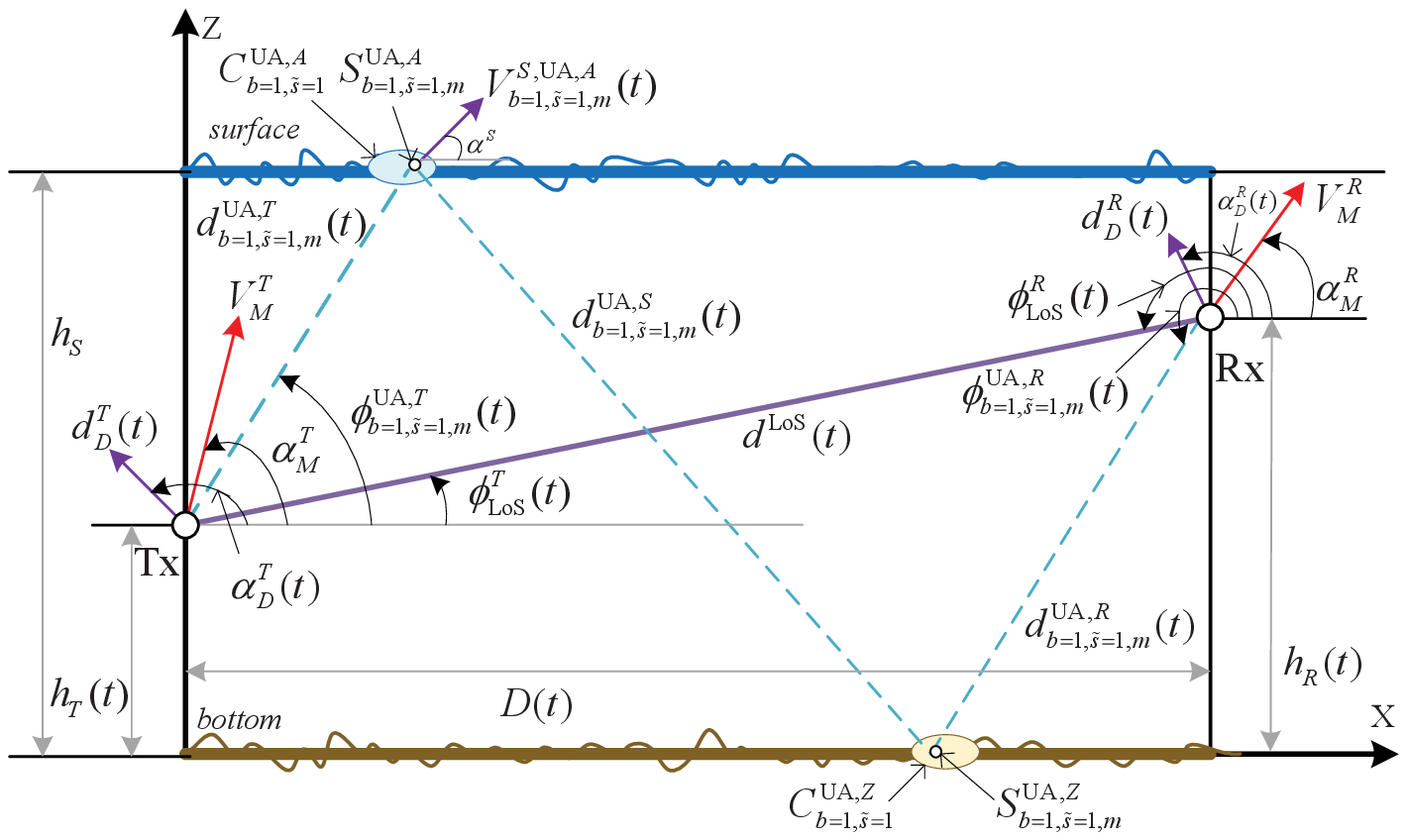}}
	\caption{A modified non-stationary shallow water GBSM.}
	\label{fig_2}
\end{figure}

Fig. \ref{fig_2} shows the proposed modified GBSM.
For clarity, only the LoS path and the $m$th ray in the UA path ($b=1$, $\tilde{s}=1$) are illustrated.
Considering a MB propagation, the AoDs of rays are only related to the first-bounce cluster $C_{b \tilde{s}}^{{\rm UA},A}$ while the AoAs of rays are only related to the last-bounce cluster $C_{b \tilde{s}}^{{\rm UA},Z}$\cite{bian2021}.
When $b+\tilde{s}=1$ is satisfied, the MB propagation is reduced to SB propagation.
The intentional motion velocity of Tx (Rx) is assumed to be constant, with speed $V_M^T$ ($V_M^R$) and travel angle $\alpha_M^T$ ($\alpha_M^R$).
The random drifting movement of Tx (Rx) is described by the displacement $d_D^T(t)$ ($d_D^R(t)$) and travel angle $\alpha_D^T(t)$ ($\alpha_D^R(t)$).
The $m$th scatterer on the surface has speed $V_{b \tilde{s}, m}^{S, {\rm UA}, A}(t)$ ($V_{b \tilde{s}, m}^{S, {\rm UA}, Z}(t)$) with constant travel angle $\alpha^S$.
The water depth, Tx depth, Rx depth, and the horizontal distance between Tx and Rx are denoted by $h_S$, $h_T(t)$, $h_R(t)$, and $D(t)$, respectively.
The transmission distance, AoD, and AoA of the LoS path are denoted by $d^{\rm LoS}(t)$, $\phi_{\rm LoS}^T(t)$ and $\phi_{\rm LoS}^R(t)$, respectively.
The AoD and AoA of the $m$th ray in the UA path are denoted by $\phi_{b \tilde{s}, m}^{{\rm UA}, T}(t)$ and $\phi_{b \tilde{s}, m}^{{\rm UA}, R}(t)$, respectively.
The transmission distances Tx$-S_{b \tilde{s}, m}^{{\rm UA}, A}$, $S_{b \tilde{s}, m}^{{\rm UA}, A}-S_{b \tilde{s}, m}^{{\rm UA}, Z}$, and $S_{b \tilde{s}, m}^{{\rm UA}, Z}-$Rx are denoted by $d_{b \tilde{s}, m}^{{\rm UA}, T}(t)$, $d_{b \tilde{s}, m}^{{\rm UA}, S}(t)$, and $d_{b \tilde{s}, m}^{{\rm UA}, R}(t)$, respectively.
For DA paths, there are similar denotations.

\subsection{Channel Transfer Function (CTF)}
Based on the proposed GBSM shown in Fig. \ref{fig_2}, the CTF of the channel model is the superposition of LoS, DA, and UA components,
which can be expressed as
\begin{equation}\label{eq_CTF}
	\begin{aligned}
		H(t, f) & =\sqrt{\frac{K}{K+1}} H^{\rm LoS}(t, f)                                                              \\
		        & +\sqrt{\frac{\eta_{\rm DA}}{K+1}} H^{\rm DA}(t, f)+\sqrt{\frac{\eta_{\rm UA}}{K+1}} H^{\rm UA}(t, f)
	\end{aligned}
\end{equation}
where $K$ is the Rice factor,
$\eta_{\rm DA}$ ($\eta_{\rm UA}$) denotes the ratio of DA (UA) component's power to the total non-LoS (NLoS) power, with constraint $\eta_{\rm DA}+\eta_{\rm UA}=1$.
The three propagation components in (\ref{eq_CTF}) are modeled as
\begin{subequations}\label{eq_CTF_3_parts}
	\begin{equation}
		H^{\rm LoS}(t, f)=a^{\rm LoS}(t, f) e^{-j 2 \pi\left(f_{c}+f\right) \tau^{\rm LoS}(t)}
	\end{equation}
	\begin{equation}
		\begin{aligned}
			H^{\rm DA}(t, f)= & \frac{1}{\sqrt{2 N_{S} N_{s \tilde{b}}^{\rm DA}}} \sum\nolimits_{s=1}^{N_{S}} \sum\nolimits_{\tilde{b}=s-1}^{s} \sum\nolimits_{n=1}^{N_{s \tilde{b}}^{\rm DA}} \\
			                  & a_{s \tilde{b}, n}^{\rm DA}(t, f) e^{j \theta_{s \tilde{b}, n}^{\rm DA}-j 2 \pi\left(f_{c}+f\right) \tau_{s \tilde{b}, n}^{\rm DA}(t)}
		\end{aligned}
	\end{equation}
	\begin{equation}
		\begin{aligned}
			H^{\rm UA}(t, f)= & \frac{1}{\sqrt{2 N_{B} M_{b \tilde{s}}^{\rm UA}}} \sum\nolimits_{b=1}^{N_{B}} \sum\nolimits_{\tilde{s}=b-1}^{b} \sum\nolimits_{m=1}^{M_{b \tilde{s}}^{\rm UA}} \\
			                  & a_{b \tilde{s}, m}^{\rm UA}(t, f) e^{j \theta_{b \tilde{s}, m}^{\rm UA}-j 2 \pi\left(f_{c}+f\right) \tau_{b \tilde{s}, m}^{\rm UA}(t)}.
		\end{aligned}
	\end{equation}
\end{subequations}
Here, $f_c$ is the carrier frequency, $a^{\rm LoS}(t, f)$ and $\tau^{\rm LoS}(t)$ are path gain and transmission delay of the LoS path, respectively.
Parameters $a_{s \tilde{b}, n}^{\rm DA}(t, f)$ ($a_{b \tilde{s}, m}^{\rm UA}(t, f)$), $\tau_{s \tilde{b}, n}^{\rm DA}(t)$ ( $\tau_{b \tilde{s}, m}^{\rm UA}(t)$), and $\theta_{s \tilde{b}, n}^{\rm DA}$ ($\theta_{b \tilde{s}, m}^{\rm UA}$) are the gain, delay, and initial phase shift of the $n$th ($m$th) ray in the DA (UA) path, respectively.
The initial phase shifts are modeled as independent random variables with uniform distributions over $[0,2\pi)$.
Derivations of the rest model parameters are given in the remainder of this section.

\subsection{Time-varying Delays}
Due to multiple motion factors in UWA channels, transmission delays are changing with time
and the motion factors can be divided into three types\cite{qarabaqi2013,baktash2015}:
1) intentional platform motion, e.g., AUV's vehicular motion; 2) unintentional drifting platform motion;
3) surface motion.
The speeds $V_M^X$ and directions of travel $\alpha_M^X$ of intentional motion are often given in measurement campaigns and are assumed to be constant here.
For the drifting motion, considering the randomness of the water movement, it is assumed that the drifting velocity vectors $\vec{V}_{D}^{X}(t)$ remain unchanged for a period of time,
and change at intervals with the speeds randomly generated between $[v_{D}^{\rm min},v_{D}^{\rm max}]$ and the angles of travel randomly generated between $[0,2\pi)$.
The change frequency is denoted by $f_v^D$.
Then the lengths of displacements and directions of drifting are calculated as $d_D^X(t)=\|\int \vec{V}_D^X(t) dt\|$ and $\alpha_D^X(t)=\arg\{\int \vec{V}_D^X(t) dt\}$, respectively.
Here, $X=\{T,R\}$, where the superscripts $T$ and $R$ denote the Tx and Rx, respectively.
For the surface motion, the relative motion speed of the $k$th scatterer on the surface to the corresponding cluster is modeled as $V_k^S(t)=2\pi f^S A^S {\cos} (2\pi f^S t+\theta_k^S)$, with a constant angle of travel $\alpha^S$.
Similar assumption can be found in \cite{qarabaqi2013}.
The phase term $\theta_k^S$ is randomly generated between $[0,2\pi)$, characterizing the randomness of the initial periodic motion of each scatterer on the surface.

\subsubsection{LoS}
For the LoS path, the time-varying delay is calculated as $\tau^{\rm LoS}(t)=d^{\rm LoS}(t)/c$, where $c$ denotes the sound speed in the water. The time-varying distance is given by
\begin{equation}\label{eq_dis_LoS}
	\begin{aligned}
		d^{\rm LoS}(t) \approx & \sqrt{D^{2}(t)+\left[h_{R}(t)-h_{T}(t)\right]^{2}}                       \\
		                       & -d_{D}^{T}(t) \cos \left[\alpha_{D}^{T}(t)-\phi_{\rm LoS}^{T}(t)\right]  \\
		                       & -d_{D}^{R}(t) \cos \left[\phi_{\rm LoS}^{R}(t)-\alpha_{D}^{R}(t)\right].
	\end{aligned}
\end{equation}
Note that the impacts of drifting motion on the positions of Tx and Rx are small, so only the influence of intentional motion on the channel geometry is considered.
Denoting the initial distances as $D(t_0)$, $h_T(t_0)$, and $h_R(t_0)$, the time-varying distances in (\ref{eq_dis_LoS}) are decided by
\begin{subequations}\label{eq_tv_distances}
	\begin{equation}
		D(t)=D\left(t_{0}\right)-V_{M}^{T} t \cdot \cos \alpha_{M}^{T}+V_{M}^{R} t \cdot \cos \alpha_{M}^{R}
	\end{equation}
	\begin{equation}
		h_{T}(t)=h_{T}\left(t_{0}\right)+V_{M}^{T} t \cdot \sin \alpha_{M}^{T}
	\end{equation}
	\begin{equation}
		h_{R}(t)=h_{R}\left(t_{0}\right)+V_{M}^{R} t \cdot \sin \alpha_{M}^{R}.
	\end{equation}
\end{subequations}
The AoD and AoA of LoS path in (\ref{eq_dis_LoS}) can be calculated as $\phi_{\rm Los}^{T}(t)=\arctan \frac{h_{R}(t)-h_{T}(t)}{D(t)}$ and $\phi_{\rm LoS}^{R}(t)=\phi_{\rm LoS}^{T}(t)+\pi$, respectively.

\subsubsection{NLoS (DA \& UA)}
Based on geometric relationships, the transmission distances and characteristic angles of incidence (AOIs) of macro-scattering rays are determined by
\begin{align}
	d_{s \tilde{b}}^{\rm DA}(t)       & =\sqrt{D^{2}(t)+\left[2 s h_{S}+(-1)^{s-\tilde{b}} h_{T}(t)-h_{R}(t)\right]^{2}}         \\
	d_{b \tilde{s}}^{\rm UA}(t)       & =\sqrt{D^{2}(t)+\left[2 \tilde{s} h_{S}-(-1)^{b-\tilde{s}} h_{T}(t)+h_{R}(t)\right]^{2}} \\
	\varphi_{s \tilde{b}}^{\rm DA}(t) & =\arctan \frac{D(t)}{2 s h_{S}+(-1)^{s-\tilde{b}} h_{T}(t)-h_{R}(t)}                     \\
	\varphi_{b \tilde{s}}^{\rm UA}(t) & =\arctan \frac{D(t)}{2 \tilde{s} h_{S}-(-1)^{b-\tilde{s}} h_{T}(t)+h_{R}(t)}
\end{align}
where examples of AOIs $\varphi_{s \tilde{b}}^{\rm DA}$ and $\varphi_{b \tilde{s}}^{\rm UA}$ are shown in Fig. \ref{fig_1}.

For brevity, we use simplified symbols to describe the derivations.
Considering the propagation route Tx$-S^A-S^Z-$Rx of a ray in a NLoS path, where $S^A$ ($S^Z$) denotes the scatterer in the first- (last-) bounce cluster $C^A$ ($C^Z$),
the transmission delay of the ray can be obtained by $\tau_r(t)=[d_r^T(t)+d_r^S(t)+d_r^R(t)]/c$, where $d_r^T(t)$, $d_r^S(t)$, and $d_r^R(t)$ denote $d({\rm Tx}-S^A)$, $d(S^A-S^Z)$, and $d(S^Z-{\rm Rx})$, respectively.
The AoD and AoA of the micro-scattering ray are denoted by $\phi_r^T(t)$ and $\phi_r^R(t)$, respectively.
Similarly, $d_c^T(t)$, $d_c^S(t)$, and $d_c^R(t)$ denote $d({\rm Tx}-C^A)$, $d(C^A-C^Z)$, and $d(C^Z-{\rm Rx})$, respectively.
The transmission distance, AoD, AoA, and characteristic AOI of the macro-scattering ray are denoted by $d_c(t)$, $\phi_c^T(t)$, $\phi_c^R(t)$, and $\varphi_c(t)$, respectively.
Note that the relevant distances need to be classified and calculated separately according to different interactions of rays with the surface and the bottom.
For DA (UA) paths, $C^Z$s are located on the surface (bottom), $C^A$s are on the bottom (surface) if $s=\tilde{b}$ ($b=\tilde{s}$) and on the surface (bottom) if not.

For MB propagation (DA paths: $s+\tilde{b} \neq 1$, UA paths: $b+\tilde{s} \neq 1$), the first and last distance can be derived as
\begin{subequations}\label{eq_distances_micro_ray}
	\begin{equation}
		d_{r}^{T}(t) \approx\left\{\begin{array}{l}
			A_{r}^{T}(t)+\frac{h_{S}-h_{T}(t)}{\sin \phi_{r}^{T}(t)}-B_{r}^{T}(t), C^{A} \text { at surface } \\
			\frac{h_{T}(t)}{\sin \left[2 \pi-\phi_{r}^{T}(t)\right]}-B_{r}^{T}(t), C^{A} \text { at bottom }
		\end{array}\right.
	\end{equation}
	\begin{equation}
		d_{r}^{R}(t) \approx\left\{\begin{array}{l}
			A_{r}^{R}(t)+\frac{h_{S}-h_{R}(t)}{\sin \left[\pi-\phi_{r}^{R}(t)\right]}-B_{r}^{R}(t), C^{Z} \text { at surface } \\
			\frac{h_{R}(t)}{\sin \left[\phi_{r}^{R}(t)-\pi\right]}-B_{r}^{R}(t), C^{Z} \text { at bottom}.
		\end{array}\right.
	\end{equation}
\end{subequations}
In (\ref{eq_distances_micro_ray}), $A_r^X(t)=A^{S} \sin \left(2 \pi f^{S} t+\theta_r^{S,X}\right) \cos \left[\phi_r^X(t)-\alpha^{S}\right]$, $B_r^X(t)=d_{D}^{X}(t) \cos \left[\alpha_{D}^{X}(t)-\phi_r^X(t)\right]$,
$\phi_r^X(t)$ are assumed to have Gaussian distributions, and the mean AoA and AoD are determined by $\phi_c^X(t)$, where $X=\{T,R\}$.
The $\theta_r^{S,T}$ ($\theta_r^{S,R}$) denotes the phase of sinusoidal speed of the surface scatterer in $C^A$ ($C^Z$).
Two parameters are used to control the angle spreads,
i.e., $\phi_r^X(t) \sim N(\phi_c^X(t),\sigma_{\phi,s}^2)$ for surface-interacting rays and $\phi_r^X(t) \sim N(\phi_c^X(t),\sigma_{\phi,b}^2$) for bottom-interacting rays.
For $C^A$ ($C^Z$), $\phi_c^T(t)=\pi/2-\varphi_c(t)$ ($\phi_c^R(t)=\pi/2+\varphi_c(t)$) and $\phi_c^T(t)=3\pi/2+\varphi_c(t)$ ($\phi_c^R(t)=3\pi/2-\varphi_c(t)$) when interacting with surface and bottom, respectively.

The second distance of the micro-scattering ray is assumed to be very close to the distance of the macro-scattering ray, and is modeled as $d_r^S(t)=d_c^S(t) \cdot e^{\Delta d^S}$,
where the $\Delta d^S$ is a zero-mean Gaussian random variable with variance $\sigma_{ds}^2$.
Parameter $\sigma_{ds}$ is used to characterize small differences between the second distances of micro-scattering rays and the macro-scattering ray and it is set as a small value of 0.001 arbitrarily in this paper.
Then there is $d_c^S(t)=d_c(t)-d_c^T(t)-d_c^R(t)$, where $d_c^X(t)=H(t)/\cos(\varphi_c(t)$, $X=\{T,R\}$.
For surface-interacting (bottom-interacting) clusters, $H(t)=h_S-h_X(t)$ ($H(t)=h_X(t)$), where $X=T$ for $C^A$ and $X=R$ for $C^Z$.

For SB propagation (DA path: $s=1,\tilde{b}=0$, UA path: $b=1,\tilde{s}=0$)
the angles are correlated in terms of geometric relationships and the second distance is zero, i.e., $d_r^S(t)=0$.
The AoDs of DA and UA paths in SB propagation can be determined by
$\phi_r^T(t)=\arctan \frac{h_{S}-h_{T}(t)}{D(t)-\left[h_{S}-h_{R}(t)\right] / \tan \left[\pi-\phi_r^R(t)\right]}$ and
$\phi_r^T(t)=2 \pi-\arctan \frac{h_{T}(t)}{D(t)-h_{R}(t) / \tan \left[\phi_r^R(t)-\pi\right]}$, respectively,
where $\theta_r^{S,T}=\theta_r^{S,R}$ for DA path.

\subsection{Time-frequency Varying Gains}
\subsubsection{LoS}
The gain of the LoS path can be expressed as
\begin{equation}\label{eq_gain_LoS}
	a^{\rm LoS}(t, f)=L_{s}\left(d^{\rm LoS}(t)\right) L_{a}\left(d^{\rm LoS}(t), f\right)
\end{equation}
where the geometric spreading loss coefficient at distance $d$~(m) is $L_{s}(d)=1/d$ \cite{zajic2011}
as we assume that Tx is a point source approximately which generates spherical spreading.
The absorption loss coefficient in (\ref{eq_gain_LoS}) can be expressed as $L_{a}(d, f)=10^{-\frac{d \cdot \alpha(f)}{20000}}$ \cite{zajic2011},
where $f$ is the frequency (kHz), $\alpha(f)$ is the frequency-dependent absorption parameter (dB/km) which can be given by the empirical Thorp model
$\alpha(f)=\frac{0.11f^{2}}{1+f^{2}}+\frac{44f^{2}}{4100+f^{2}}+2.75 \times 10^{-4} f^{2}+0.003$\cite{brekhovskikh2003}.

\subsubsection{NLoS (DA \& UA)}
The gains of DA (UA) paths are related to the transmission distances, AOIs and frequency.
As the distances and angles of micro-scattering rays are very close to that of the corresponding macro-scattering ray, we assume the gains of micro-scattering rays in a NLoS path are equal to that of the macro-scattering ray,
i.e., $a_{s \tilde{b}, n}^{\rm DA}(t, f)=a_{s \tilde{b}}^{\rm DA}(t, f)$ and $a_{b \tilde{s}, m}^{\rm UA}(t, f)=a_{b \tilde{s}}^{\rm UA}(t, f)$ for the sake of simplicity.
The gains can be expressed as
\begin{align}
	a_{s \tilde{b}}^{\rm DA}(t, f) & =L_{s}\left(d_{s \tilde{b}}^{\rm DA}(t)\right) L_{a}\left(d_{s \tilde{b}}^{\rm DA}(t), f\right) L_{b}\left(\varphi_{s \tilde{b}}^{\rm DA}(t)\right)^{\tilde{b}} \\
	a_{b \tilde{s}}^{\rm UA}(t, f) & =L_{s}\left(d_{b \tilde{s}}^{\rm UA}(t)\right) L_{a}\left(d_{b \tilde{s}}^{\rm UA}(t), f\right) L_{b}\left(\varphi_{b \tilde{s}}^{\rm UA}(t)\right)^{b}
\end{align}
where $L_b(\cdot)$ is the bottom reflection loss and is given by \cite{brekhovskikh2003}
\begin{equation}\label{eq_bottom_loss}
	L_{b}(\varphi)=\left| \frac{\left(\rho_{b} / \rho_{w}\right) \cos (\varphi)-\sqrt{\left(c_{w} / c_{b}\right)^{2}-\sin ^{2}(\varphi)}}{\left(\rho_{b} / \rho_{w}\right) \cos (\varphi)+\sqrt{\left(c_{w} / c_{b}\right)^{2}-\sin ^{2}(\varphi)}}\right|
\end{equation}
where $\varphi$ is the AoI at the bottom, $\rho_b$ ($\rho_w$) denotes the density of the bottom (water), and $c_b$ ($c_w$) denotes the sound speed in the bottom (water).

\section{Statistical properties}
\subsection{Time-Frequency Correlation Function (TF-CF)}
The TF-CF of the channel model can be defined as
\begin{equation}\label{eq_TFCF}
	R_{H}(t, f ; \Delta t, \Delta f)=\mathrm{E}\left\{H(t, f) H^{*}(t-\Delta t, f-\Delta f)\right\}
\end{equation}
where $\mathrm{E}\{\cdot\}$ denotes the ensemble average and $(\cdot)^{*}$ is the complex conjugate operation.
Assuming that LoS, DA, and UA paths are independent with each other,
by substituting (\ref{eq_CTF}) and (\ref{eq_CTF_3_parts}) into (\ref{eq_TFCF}), the TF-CF can be expressed as
\begin{equation}\label{eq_TFCF_2}
	\begin{array}{l}
		R_{H}(t, f ; \Delta t, \Delta f)=\frac{K}{K+1} R_{H}^{\rm LoS}(t, f ; \Delta t, \Delta f)                                                   \\
		\quad+\frac{\eta_{\rm DA}}{2 N_{S}(K+1)} \sum_{s=1}^{N_{S}} \sum_{\tilde{b}=s-1}^{s} R_{H, s \tilde{b}}^{\rm DA}(t, f ; \Delta t, \Delta f) \\
		\quad+\frac{\eta_{\rm UA}}{2 N_{B}(K+1)} \sum_{b=1}^{N_{B}} \sum_{\tilde{s}=b-1}^{b} R_{H, b \tilde{s}}^{\rm UA}(t, f ; \Delta t, \Delta f).
	\end{array}
\end{equation}
\begin{figure*}[t]
	\begin{subequations}\label{eq_TFCF_3_parts}
		\begin{align}
			R_{H}^{\rm LoS}(t, f ; \Delta t, \Delta f)             & =\mathrm{E}\left\{A^{\rm LoS} e^{-j 2 \pi\left(f_{c}+f\right)\left[\tau^{\rm LoS}(t)-\tau^{\rm LoS}(t-\Delta t)\right]-j 2 \pi \Delta f \tau^{\rm LoS}(t-\Delta t)}\right\}                                                                                                                                          \\
			R_{H, s \tilde{b}}^{\rm DA}(t, f ; \Delta t, \Delta f) & =\mathrm{E}\left\{\frac{1}{N_{s \tilde{b}}^{\rm DA}} \sum_{n=1}^{N_{s \tilde{b}}^{\rm DA}} A_{s \tilde{b}, n}^{\rm DA} e^{-j 2 \pi\left(f_{c}+f\right)\left[\tau_{s \tilde{b}, n}^{\rm DA}(t)-\tau_{s \tilde{b}, n}^{\rm DA}(t-\Delta t)\right]-j 2 \pi \Delta f \tau_{s \tilde{b}, n}^{\rm DA}(t-\Delta t)}\right\} \\
			R_{H, b \tilde{s}}^{\rm UA}(t, f ; \Delta t, \Delta f) & =\mathrm{E}\left\{\frac{1}{M_{b \tilde{s}}^{\rm UA}} \sum_{m=1}^{M_{b \tilde{s}}^{\rm UA}} A_{b \tilde{s}, m}^{\rm UA} e^{-j 2 \pi\left(f_{c}+f\right)\left[\tau_{b \tilde{s}, m}^{\rm UA}(t)-\tau_{b \tilde{s}, m}^{\rm UA}(t-\Delta t)\right]-j 2 \pi \Delta f \tau_{b \tilde{s}, m}^{\rm UA}(t-\Delta t)}\right\}
		\end{align}
	\end{subequations}
	{\noindent} \rule[-10pt]{18.07cm}{0.1em}
\end{figure*}
Denoting that $A^{\rm LoS}=a^{\rm LoS}(t, f) a^{\rm LoS}(t-\Delta t, f-\Delta f)$, $A_{s \tilde{b}, n}^{\rm DA}=a_{s \tilde{b}, n}^{\rm DA}(t, f) a_{s \tilde{b}, n}^{\rm DA}(t-\Delta t, f-\Delta f)$, and $A_{b \tilde{s}, m}^{\rm UA}=a_{b \tilde{s}, m}^{\rm UA}(t, f) a_{b \tilde{s}, m}^{\rm UA}(t-\Delta t, f-\Delta f)$,
the correlation functions of LoS, DA, and UA components are given by (\ref{eq_TFCF_3_parts}a)$-$(\ref{eq_TFCF_3_parts}c), shown at the top of the next page.
The temporal ACF can be obtained by the TF-CF, i.e., $R_H^{\rm ACF}(t,f;\Delta t)=R_H(t,f;\Delta t, 0)$.

\subsection{PDP}
The time-frequency varying PDP can be expressed as
\begin{equation}\label{eq_PDP}
	\begin{array}{l}
		P(t, f ; \tau)=\frac{K}{K+1} P_{\rm LoS}(t, f ; \tau)                                                                                                                                   \\
		+\frac{\eta_{\rm DA}}{2 N_{S} N_{s \tilde{b}}^{\rm DA}(K+1)} \sum_{s=1}^{N_{S}} \sum_{\tilde{b}=s-1}^{s} \sum_{n=1}^{N_{s \tilde{b}}^{\rm DA}} P_{s \tilde{b}, n}^{\rm DA}(t, f ; \tau) \\
		+\frac{\eta_{\rm UA}}{2 N_{B} M_{b \tilde{s}}^{\rm UA}(K+1)} \sum_{b=1}^{N_{B}} \sum_{\tilde{s}=b-1}^{b} \sum_{m=1}^{M_{b \tilde{s}}^{\rm UA}} P_{b \tilde{s}, m}^{\rm UA}(t, f ; \tau).
	\end{array}
\end{equation}
In (\ref{eq_PDP}), the PDPs of three propagation components are determined as {\small $P_{\rm LoS}(t, f ; \tau)=\left[a_{\rm LoS}(t, f)\right]^{2} \delta\left(\tau-\tau_{\rm LoS}(t)\right)$}, {\small $P_{s \tilde{b}, n}^{\rm DA}(t, f ; \tau)=\left[a_{s \tilde{b}, n}^{\rm DA}(t, f)\right]^{2} \delta\left(\tau-\tau_{s \tilde{b}, n}^{\rm DA}(t)\right)$}, and
	{\small $P_{b \tilde{s}, m}^{\rm UA}(t, f ; \tau)=\left[a_{b \tilde{s}, m}^{\rm UA}(t, f)\right]^{2} \delta\left(\tau-\tau_{b \tilde{s}, m}^{\rm UA}(t)\right)$}, respectively.

\subsection{Average Delay and RMS Delay Spread}
The average delay and the RMS delay spread can be obtained by the PDP, and can be calculated as
\begin{equation}\label{eq_ave_delay}
	\mu_{\tau}(t, f)=\frac{\sum_{\tau} \tau \cdot P(t, f ; \tau)}{\sum_{\tau} P(t, f ; \tau)}
\end{equation}
\begin{equation}\label{eq_delay_spread}
	\sigma_{\tau}(t, f)=\sqrt{\frac{\sum_{\tau}\left(\tau-\mu_{\tau}(t, f)\right)^{2} \cdot P(t, f ; \tau)}{\sum_{\tau} P(t, f ; \tau)}}.
\end{equation}

\section{Results and Analysis}
In this section, some key statistical properties of the proposed model are simulated and verified with the corresponding measurement data.
In the simulation, we consider a scenario with geometry parameters given as $D(t_0)=2000$~m, $h_S=100$~m, $h_T(t_0)=50$~m, and $h_R(t_0)=80$~m.
The sound speed $c (c_w)$ is set as $1500$~m/s \cite{stojanovic2009}.
We assume the drifting motion changes once per second and consider vertical surface motion, i.e., $f_v^D=1$~Hz and $\alpha^S=\pi /2$.
Since shallow UWA communication usually accounts for long-range communication, the transmission distance is usually large compared with the water depth.
Thus, we assume that the angle spread is small and set the corresponding parameter as $\sigma_{\phi,s}=\sigma_{\phi,b}=0.015$.
And refer to \cite{naderi2017}, we set $N_B=N_S=2$, $\eta_{\rm DA}=\eta_{\rm UA}=0.5$, $\rho_b/\rho_w=1.5$, and $c_b=1600$~m/s.

\begin{figure}[tb]
	\centerline{\includegraphics[width=0.35\textwidth]{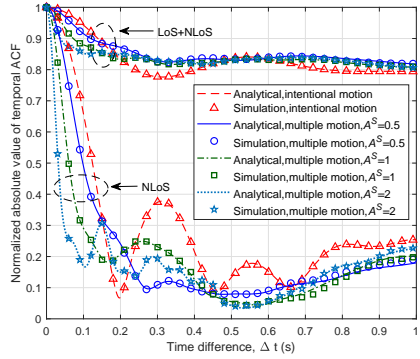}}
	\caption{The temporal ACFs of the proposed model under different motion influences with and without LoS component ($K=5$, $V_M^T=V_M^R=1$~m/s, $\alpha_M^T=0$, $\alpha_M^R=-\pi/2$, $v_{D}^{\rm min}=0.1$~m/s, $v_{D}^{\rm max}=0.12$~m/s, $f^S=0.5$~Hz, $f_c=15$~kHz).}
	\label{fig_3}
\end{figure}
Fig. \ref{fig_3} illustrates the normalized absolute values of temporal ACFs of the proposed model under different motion influences with and without LoS component.
It can be observed that multiple motion factors in UWA channels make the temporal ACFs decay faster.
The temporal correlation remains higher and is less affected by motion factors with the LoS component.
It can also be noticed that as the amplitude of surface scatterers' periodic motion ($A^S$) increases, the ACFs decay faster.
Note that the surface motion speed is proportional to $A^S$, we can conclude that larger motion speed will attenuate the temporal correlation more dramatically.
It can also be predicted that UWA channel coherence time will be lower when ships pass through or the weather is windy.
Moreover, the analytical results have a good consistency with the simulated results, validating the correctness of the derivations and simulations.

\begin{figure}[tb]
	\subfigure[]{\centerline{\includegraphics[width=0.35\textwidth]{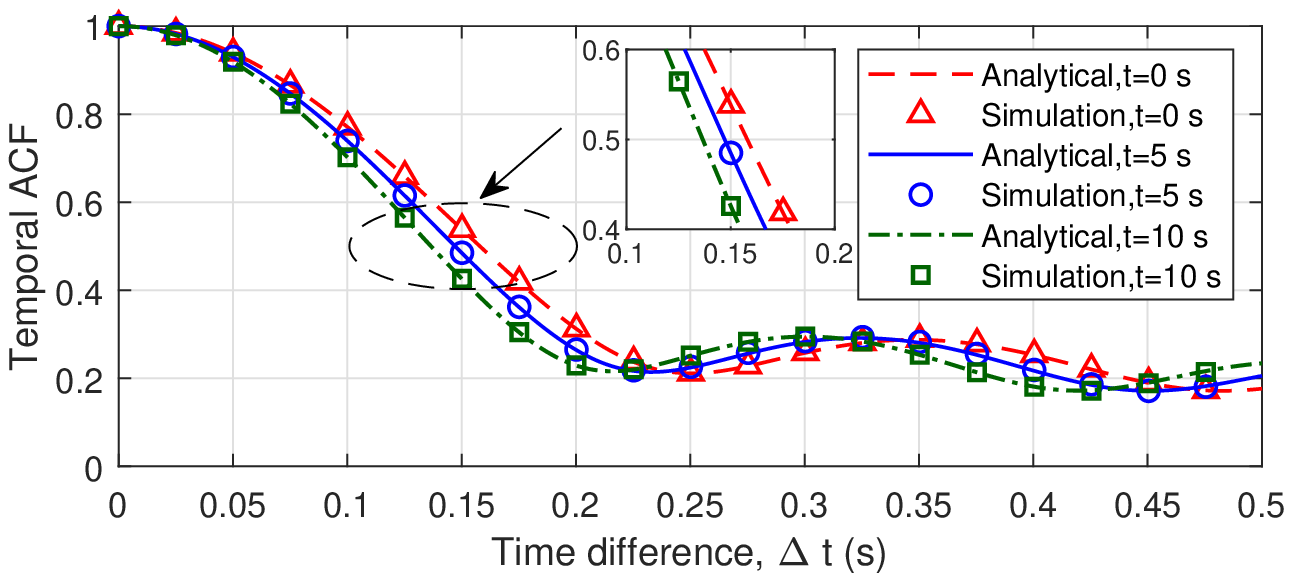}}}
	\subfigure[]{\centerline{\includegraphics[width=0.35\textwidth]{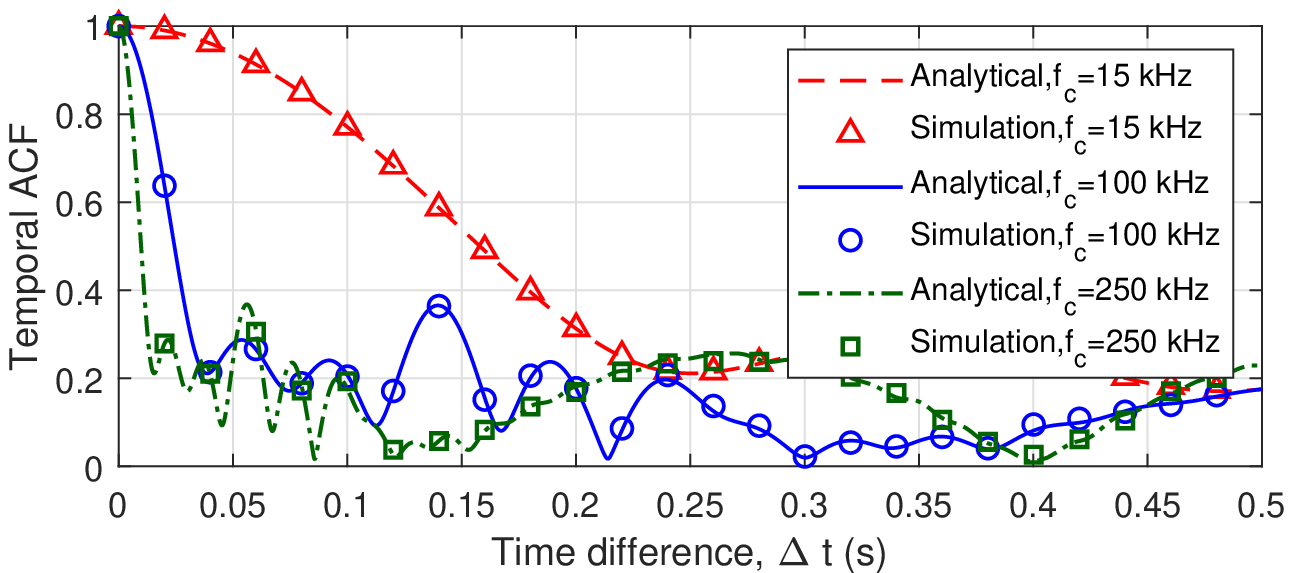}}}
	\caption{The temporal ACFs of the proposed model at different instants with different carrier frequencies ($NLoS$, $V_M^T=10$~m/s, $V_M^R=5$~m/s, $\alpha_M^T=0$, $\alpha_M^R=-\pi$, $v_{D}^{\rm min}=v_{D}^{\rm max}=0$~m/s, $A^S=0$, $f^S=0$~Hz, $f_c=15$~kHz for upper subfigure, $t=0$~s for lower subfigure).}
	\label{fig_4}
\end{figure}
Fig. \ref{fig_4} presents the temporal ACFs at different time instants and different carrier frequencies.
We can observe different temporal ACFs at $t=0$~s, $5$~s, and $10$~s,
which are resulted from time-varying angles, distances, and clusters' locations in the proposed model, illustrating that our model can mimic the non-stationarity of UWA channels.
Besides, the distinction between temporal ACFs at different carrier frequencies shows that the magnitude of the coherence time is related to the communication frequency band.
The higher the carrier frequency, the shorter the coherence time.
In addition, the simulated results of ACFs can fit the analytical results well.

\begin{figure}[tb]
	\subfigure[]{\centerline{\includegraphics[width=0.35\textwidth]{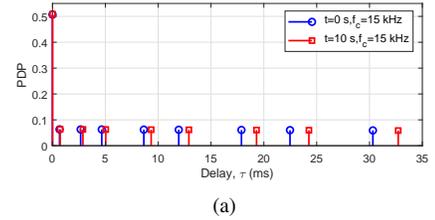}}}
	\subfigure[]{\centerline{\includegraphics[width=0.35\textwidth]{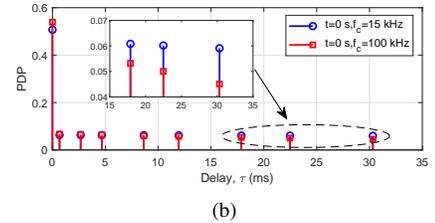}}}
	\caption{The PDPs of the modified non-stationary model at different time instants with different carrier frequencies ($K=1$, $V_M^T=10$~m/s, $V_M^R=5$~m/s, $\alpha_M^T=0$, $\alpha_M^R=-\pi$, $v_{D}^{\rm min}=v_{D}^{\rm max}=0$~m/s, $A^S=0$, $f^S=0$~Hz).}
	\label{fig_5}
\end{figure}
Fig. \ref{fig_5} shows the PDPs of the proposed model by using the mean gains and delays of rays within clusters.
All transmission delays are normalized with respect to the delay of the first arrival ray.
We can see that the PDPs are different at $t=0$~s and $5$~s, and have differences at $f_c=15$~kHz and $100$~kHz, which shows the non-stationarity of the proposed model again.
The strongest ray is the first arrival ray in the isovelocity shallow water scenarios as can be seen in Fig. \ref{fig_5}.

\begin{table}[t]
	\caption{Comparison of Delay Statistics \protect\\ ($K=1.44$, $\sigma_{\phi,s}=\sigma_{\phi,b}=0.02317$, $\eta_{\rm DA}=\eta_{\rm UA}=0.5$, $N_S=N_B=1$).}
	\begin{center}
		\begin{tabular}{|l|c|c|c|c}
			\hline
			Delay Statistics & \makecell{Measurement                           \\ data in \cite{naderi2017}}&\makecell{Simulation\\ model in \cite{naderi2017}}&\makecell{The proposed\\ model}\\
			\hline
			Average Delay    & $1.5$ ms              & $1.491$ ms & $1.505$ ms \\
			\hline
			RMS Delay Spread & $2.4$ ms              & $2.410$ ms & $2.399$ ms \\
			\hline
		\end{tabular}
		\label{tab2}
	\end{center}
\end{table}

Table \ref{tab2} gives the comparison of delay statistics of the simulation model in \cite{naderi2017} and the proposed model with the corresponding measurement data in \cite{naderi2017}.
The parameters of measurement campaign are given in as $D(t_0)=1500$~m, $h_S=80$~m, $h_T(t_0) \approx 34.5$~m, $h_R(t_0) \approx 36$~m, $c=1440$~m/s, $V_M^T=V_M^R=v_{D}^{\rm min}=v_{D}^{\rm max}=0$~m/s, $f_c=17$~kHz, $c_b/c_w=1.11$, and $\rho_b/\rho_w=1.5$ \cite{naderi2017} .
The weather of the measurement scenario in \cite{naderi2017} was rainy and windy, so we set parameters of surface motion as $A^S=2$, $f^S=0.1$~Hz, and $\alpha^S=\pi/2$ \cite{qarabaqi2013}.
With the rest of parameters chosen according to the estimation procedure introduced in \cite{wu2017general} which is based on the minimum mean square error criterion,
the proposed model matches better with measurement data, showing the practicality of our model.

\section{Conclusions}
In this paper, a modified non-stationary GBSM for shallow water UWA communication systems has been proposed.
Some important statistical properties have been investigated, and the analytical results have a good consistency with the simulated results, validating the correctness of the derivations and simulations.
Simulation results have illustrated that multiple motion factors have great influence on temporal ACFs, so as on UWA communication systems.
The impact of unintentional motion factors cannot be ignored.
The fact that simulation results of the proposed model vary with time and frequency has shown that our model can mimic the non-stationarity of UWA channels.
In addition, simulation results of statistical properties fit well with measurements, illustrating the usefulness of the proposed model.
In our future work, we will try to consider the change of sound speed to improve the applicability of the channel model in deep waters.



\section*{Acknowledgment}
\small
{This work was supported by the National Key R\&D Program of China under Grant 2018YFB1801101,
	the National Natural Science Foundation of China (NSFC) under Grant 61960206006,
	the Frontiers Science Center for Mobile Information Communication and Security,
	the High Level Innovation and Entrepreneurial Research Team Program in Jiangsu,
	the High Level Innovation and Entrepreneurial Talent Introduction Program in Jiangsu,
	the Research Fund of National Mobile Communications Research Laboratory, Southeast University,
	under Grant 2020B01, the Fundamental Research Funds for the Central Universities under Grant 2242020R30001,
	and the EU H2020 RISE TESTBED2 project under Grant 872172.}

\bibliographystyle{IEEEtran}

\begin{thebibliography}{99}

	\bibitem{wang2020_6g}
	C.-X. Wang, J.~Huang, H.~Wang, X.~Gao, X.-H. You, and Y.~Hao, ``6G wireless
	channel measurements and models: Trends and challenges,'' \emph{IEEE Veh. Technol. Mag.},
	vol.~15, no.~4, pp.~22--32, Dec. 2020.

	\bibitem{you2020_6g}
	X.-H. You, C.-X. Wang, J.~Huang, \emph{et al.}, ``Towards 6G wireless communication
	networks: Vision, enabling technologies, and new paradigm shifts,'' \emph{ Sci. China Inf. Sci.},
	vol.~64, no.~1, Jan. 2021.

	\bibitem{wang2018survey}
	C.-X. Wang, J.~Bian, J.~Sun, W.~Zhang, and M.~Zhang, ``A survey of 5G channel
	measurements and models,'' \emph{IEEE Commun. Surveys Tuts.}, vol.~20, no.~4,
	pp. 3142--3168, 4th Quart., 2018.

	\bibitem{qarabaqi2013}
	P.~Qarabaqi and M.~Stojanovic, ``Statistical characterization and
	computationally efficient modeling of a class of underwater acoustic
	communication channels,'' \emph{IEEE J. Oceanic Eng.}, vol.~38, no.~4, pp.
	701--717, Oct. 2013.

	\bibitem{baktash2015}
	E.~Baktash, M.~J. Dehghani, M.~R.~F. Nasab, and M.~Karimi, ``Shallow {{water
			acoustic channel modeling based}} on {{analytical second order statistics}}
	for {{moving transmitter}}/{{receiver}},'' \emph{IEEE Trans. Signal
		Process.}, vol.~63, no.~10, pp. 2533--2545, May 2015.

	\bibitem{rudander2017}
	J.~Rudander, T.~Husøy, P.~Orten, and P.~{van Walree}, ``Shallow-water channel
	sounding for high speed acoustic communication,'' in \emph{Proc.
			{{OCEANS}}'17}, {Aberdeen, United Kingdom}: {IEEE}, June 2017, pp. 1--8.

	\bibitem{stojanovic2009}
	M.~Stojanovic and J.~Preisig, ``Underwater acoustic communication channels:
	{{Propagation}} models and statistical characterization,'' \emph{IEEE Commun.
		Mag.}, vol.~47, no.~1, pp. 84--89, Jan. 2009.

	\bibitem{vanwalree2013}
	P.~A. {van Walree}, ``Propagation and scattering effects in underwater acoustic
	communication channels,'' \emph{IEEE J. Oceanic Eng.}, vol.~38, no.~4, pp.
	614--631, Oct. 2013.

	\bibitem{vanwalree2013a}
	P.~A. {van Walree} and R.~Otnes, ``Ultrawideband underwater acoustic
	communication channels,'' \emph{IEEE J. Oceanic Eng.}, vol.~38, no.~4, pp.
	678--688, Oct. 2013.

	\bibitem{gul2017}
	S. Gul, S. S. H. Zaidi, R. Khan and A. B. Wala, ``Underwater acoustic channel modeling using BELLHOP ray tracing method,''
	in \emph{Proc. IBCAST'17}, Islamabad, Pakistan, Jan. 2017, pp. 665-670.

	\bibitem{huang2019}
	J. Huang and R. Diamant, ``Pre-setting of channel types for long range underwater acoustic communications,''
	in \emph{Proc. OCEANS'19}, Marseille, France, Jun. 2019, pp. 1–6.

	\bibitem{socheleau2010}
	F.-X. Socheleau, C.~Laot, and J.-M. Passerieux, ``A maximum entropy framework
	for statistical modeling of underwater acoustic communication channels,'' in
	\emph{Proc. OCEANS'10}, {Sydney, Australia}: {IEEE}, May 2010, pp. 1--7.

	\bibitem{naderi2017}
	M.~Naderi, M.~Pätzold, R.~Hicheri, and N.~Youssef, ``A {{geometry}}-{{based
			underwater acoustic channel model allowing}} for {{sloped ocean bottom
					conditions}},'' \emph{IEEE Trans. Wireless Commun.}, vol.~16, no.~4, pp.
	2394--2408, Apr. 2017.

	\bibitem{naderi2018}
	M.~Naderi, A.~G. Zaji{\'c}, and M.~Pätzold, ``A nonisovelocity geometry-based
	underwater acoustic channel model,'' \emph{IEEE Trans. Veh. Technol.},
	vol.~67, no.~4, pp. 2864--2879, Apr. 2018.

	\bibitem{zajic2011}
	A.~G. Zaji{\'c}, ``Statistical modeling of {{MIMO}} mobile-to-mobile underwater
	channels,'' \emph{IEEE Trans. Veh. Technol.}, vol.~60, no.~4, pp. 1337--1351,
	May 2011.

	\bibitem{song2019}
	A. Song, M. Stojanovic, and M. Chitre,''Editorial underwater acoustic communications:
	where we stand and what is next?,'' \emph{IEEE J. Oceanic Eng.}, vol. 44, no. 1,
	pp. 1–6, Jan. 2019.
	
	\bibitem{zhou2014}
	S. Zhou and Z. Wang, \emph{OFDM for Underwater Acoustic Communications}.
	Chichester, West Sussex, United Kingdom: Wiley, 2014.

	\bibitem{bian2021}
	J. Bian, C.-X. Wang, X. Gao, X. You, and M. Zhang, “A general 3D non-stationary wireless 
	channel model for 5G and beyond,”  \emph{IEEE Trans. Wireless Commun.}, accepted for
	 publication.

	\bibitem{brekhovskikh2003}
	L.~M. Brekhovskikh and Y.~P. Lysanov, \emph{Fundamentals of Ocean Acoustics},
	3rd~ed.\hskip 1em plus 0.5em minus 0.4em\relax {New York}: {Springer}, 2003.

	\bibitem{wu2017general}
	S.~Wu, C.-X. Wang, H.~Aggoune, M.~M. Alwakeel, and X.~You, ``A general 3D non-stationary
	5G wireless channel model,'' \emph{IEEE Trans. Commun.},
	vol.~66, no.~7, pp. 3065--3078, July 2018.
	

\end{thebibliography}

\end{document}